\documentclass[a4paper,12pt]{article}

\usepackage{hyperref}
\hypersetup{colorlinks=true,citecolor=blue,urlcolor=blue,linkcolor=black}

\usepackage{url}
\usepackage{amssymb}
\usepackage[footnotesize]{caption}
\usepackage{graphicx}
\usepackage[font=scriptsize]{subcaption}
\usepackage[dvipsnames]{xcolor}
\usepackage{mathtools}
\usepackage{braket}
\usepackage{cite}
\usepackage{multirow}
\usepackage{relsize}
\usepackage{fullpage}
\usepackage{makecell}
\usepackage{blkarray}
\usepackage{placeins}
\usepackage[utf8]{inputenc}
\usepackage{empheq}
\usepackage{changepage}

\numberwithin{equation}{section}
\newcommand{\pmatr}[1]{\begin{pmatrix} #1 \end{pmatrix}}
\newcommand{\simlt}{~\mbox{\smaller\(\lesssim\)}~}
\newcommand{\simgt}{~\mbox{\smaller\(\gtrsim\)}~}

\newcommand{\order}[1]{\mathcal{O}( #1 )}
\newcommand{\ord}{$\mathcal{O}$}

\begin{document}
\begin{titlepage}

\noindent August 2016
\hfill IPMU 16-0120

\vskip 1.5cm

\setcounter{page}{0}
\vspace*{0.7cm}

\begin{center}
{\bf\Large
Leptogenesis after Chaotic Sneutrino Inflation\\\smallskip
and the Supersymmetry Breaking Scale}
\\[12mm]
Fredrik~Bj\"{o}rkeroth$^{\star}$%
\footnote{E-mail: {\tt f.bjorkeroth@soton.ac.uk}},
Stephen~F.~King$^{\star}$%
\footnote{E-mail: \texttt{s.f.king@soton.ac.uk}},
Kai Schmitz$^{\dagger}$%
\footnote{E-mail: \texttt{kai.schmitz@mpi-hd.mpg.de}},
Tsutomu T. Yanagida$^{\ddagger}$%
\footnote{E-mail: \texttt{tsutomu.tyanagida@ipmu.jp}}
\\[-2mm]

\vspace*{1cm}
{\it
$^{\star}$
Physics and Astronomy, University of Southampton,\\
SO17 1BJ Southampton, United Kingdom

$^{\dagger}$
Max Planck Institute for Nuclear Physics (MPIK),\\ 69117 Heidelberg, Germany

$^{\ddagger}$ 
Kavli IPMU (WPI), UTIAS, The University of Tokyo, \\
Kashiwa, Chiba 277-8583, Japan
}
\end{center}

\begin{abstract}
{\noindent
We discuss resonant leptogenesis arising from the decays of two nearly-degenerate right-handed neutrinos, identified as the inflaton and stabiliser superfields in a model of chaotic sneutrino inflation.
We compare an analytical estimate of the baryon asymmetry $ \eta_B $ in the Boltzmann approximation to a numerical solution of the full density matrix equations, and find that the analytical result fails to capture the correct physics in certain regions of parameter space.
The observed baryon asymmetry can be realised for a breaking of the mass degeneracy as small as $ \order{10^{-8}} $. 
The origin of such a small mass splitting is explained by considering supersymmetry (SUSY) breaking in supergravity, which requires a constant in the superpotential of the order of the gravitino mass $ m_{3/2} $ to cancel the cosmological constant. 
This yields additional terms in the (s)neutrino mass matrices, lifting the degeneracy and linking $ \eta_B $ to the SUSY breaking scale. 
We find that achieving the correct baryon asymmetry requires a gravitino mass $ m_{3/2} \geq \order{100} $ TeV.
}
\end{abstract}
\end{titlepage}

\section{Introduction}

Inflation \cite{Guth:1980zm} has established itself as the most promising candidate for explaining the physics of the very early universe, and agrees with all observations to date. Many attempts have been made to understand inflation from a field theory perspective (for reviews, see \cite{Lyth:1998xn}), with the central question: what field plays the role of the inflaton?

In supersymmetry (SUSY), an interesting possibility presents itself wherein the inflaton is the scalar component of a right-handed (RH) neutrino superfield, so-called sneutrino inflation \cite{Murayama:1992ua}. This in turn implies a type I seesaw mechanism \cite{seesaw} giving light neutrino masses, and allows for leptogenesis \cite{Fukugita:1986hr} from the RH (s)neutrino decays. In short, we may couple two cosmological phenomena -- inflation and the baryon asymmetry of the universe (BAU) -- to low-energy neutrino data.
The observed BAU is given by \cite{1502.01589}
\begin{equation}
	\eta_B^{\mathrm{obs}} = (6.08 \pm 0.04) \times 10^{-10}.
\end{equation}

A viable model for chaotic sneutrino inflation in supergravity (SUGRA)  is developed in \cite{1601.00192}, and summarised in this paper. The seesaw mechanism arises naturally from the model, as do the conditions for leptogenesis.
In order to prevent super-Planckian masses of the leptons and Higgs, the model assumes an unbroken discrete shift symmetry in the superpotential. 
The model is shown to be compatible with the cosmological data for the spectral index $ n_s $ and the tensor-to-scalar ratio $ r $ of the primordial scalar power spectrum.

Several models for chaotic (sneutrino) inflation have been proposed in the literature with different mechanisms for preventing super-Planckian masses. 
In \cite{Antusch:2009ty}, this is achieved by introducing a Heisenberg symmetry, while the slope of the inflaton potential arises from a small Heisenberg symmetry breaking term.
More recently, it has been shown that the hyperbolic geometry of $ \alpha $-attractor models can similarly protect the lepton and Higgs masses \cite{1607.08854}.
Further works on sneutrino inflation in the literature may be found in \cite{sneutrinoinflation}.%
\footnote{For a brief literature review, we refer the reader to \cite{1601.00192}.}

The inflaton and stabiliser fields necessary for chaotic inflation in SUGRA \cite{Kawasaki:2000yn} are identified with two RH neutrino superfields $ N_{1,2} $, of degenerate masses $ M_{1,2} = M \sim 10^{13} $ GeV. The associated reheating temperature $ T_{R} $ is calculated to be $ \order{10^{14}} $ GeV. This sets the stage for resonant thermal leptogenesis \cite{Pilaftsis:1997jf}, on the condition that there is some small splitting between the neutrino masses. Typically, resonant leptogenesis is considered to be realized at low energy scales. In our scenario, we encounter by contrast the rather unusual and to some extent novel case of resonant leptogenesis realized at a high energy scale.

In this paper we examine the above scenario of resonant leptogenesis in detail, examining the conditions under which the correct BAU may be produced, taking into account the constraints from data on neutrino masses and mixing. We find that even an extremely small mass splitting, of $ \order{10^{-8}} $, can produce the correct asymmetry.

We present a compelling explanation for the origin of such a small mass splitting by considering SUSY breaking in SUGRA. Our main point is the following: in SUGRA, the superpotential must contain a constant term proportional to the gravitino mass, $ W \supset m_{3/2} M_P^2 $, so as to achieve vanishing cosmological constant in the SUSY-breaking vacuum. Such a constant may result from dynamical $R$ symmetry breaking in a hidden sector.
We find that it leads to the breaking of the mass degeneracy for both fermionic and pseudoscalar RH neutrinos, with a mass squared difference of $ \order{m_{3/2}M} $. 
This consequently links the observed value of the BAU with the SUSY breaking scale and the gravitino mass; we find that naturally $ m_{3/2} \simgt 800 $ TeV, though it may be lower by an \ord(1) factor.

The paper is organised as follows: in Section \ref{sec:model} we summarise the inflation model in \cite{1601.00192}, and define the neutrino Yukawa and Majorana mass matrices. In Section \ref{sec:approx} we derive an analytical expression for the BAU, in the Boltzmann approximation, from the decay of two heavy, nearly-degenerate neutrinos. 
However, this approximate expression is unreliable in the presence of heavy flavour effects, and the $ B-L $ asymmetry should be resolved in the full density matrix formalism. The density matrix equation is given in Section \ref{sec:exact}. In Section \ref{sec:numerics} we solve the system numerically, and plot the resultant BAU in terms of the free parameters of the theory. In Section \ref{sec:deltaM} we describe how a small splitting in the RH neutrino masses may arise as a consequence of SUSY breaking in SUGRA, and discuss the implications of an \ord(100) TeV gravitino. Section \ref{sec:conclusion} concludes.

\section{Chaotic inflation model}
\label{sec:model}
\subsection{Sneutrinos as inflaton and stabiliser fields}
We base our study of leptogenesis on an existing model of chaotic sneutrino inflation \cite{1601.00192}, although the results may be applied more generally to high-scale resonant leptogenesis.
The model consists of two singlet superfields, the inflaton ($ \Phi $) and stabiliser ($ S $) superfield.  They couple in a superpotential term like $ W \supset M \Phi S $, as well as to lepton and Higgs doublets via supersymmetric Yukawa couplings. 
We identify them as right-handed (s)neutrinos, i.\,e., $ \Phi \equiv N_1 $ and $ S \equiv N_2 $, and the (scalar) inflaton field as $ \phi \equiv \sqrt{2}\:{\rm Im}[N_1] $.
Slow-roll inflation requires that there be no additional sizable mass terms for the fields $N_{1,2}$ in the superpotential, apart from the Dirac mass term $M N_1 N_2$. 
This may be achieved by invoking a symmetry%
\footnote{An appropriate symmetry would be a global $U(1)$ or $\mathbb{Z}_n$ where $N_1$ and $N_2$ have opposite charge. The minimal discrete symmetry that forbids renormalisable RH neutrino terms other than $M N_1 N_2$ is $ \mathbb{Z}_4 $, where neutrinos have charges 1 and 3.}
in the neutrino mass sector, under which diagonal terms like $N_1^2$, $N_2^2$ are forbidden.

The observed amplitude of the primordial scalar power spectrum fixes the mass scale $M$ at $ M \sim 10^{13} $ GeV.
The relevant superpotential and K\"ahler potential are given by
\begin{align}
	K &= \frac{1}{2} (N_1 + N_1^\dagger)^2 + |N_2|^2 - k_2 \frac{|N_2|^4}{M_P^2}, 
	\label{eq:Kinf}\\
	W &= M N_1 N_2 + \tilde{h}_{\alpha i} H^u L_\alpha N_i , \label{eq:Winf}
\end{align}
where $ k_2 $ is an \ord(1) constant and we expect Yukawa couplings $ \tilde{h}_{\alpha i} \sim \order{0.1} $ so as to obtain standard model neutrino masses in the $10 - 100$ meV range.

At tree level, the K\"ahler potential respects a shift symmetry in the direction of $ \phi $, i.\,e.,
\begin{equation}
	\phi \rightarrow \phi + A \,, \quad A \in \mathbb{R} .
\end{equation}
This ensures the flatness of the inflaton potential at $ \phi > M_P $.
The shift symmetry in the K\"ahler potential is an approximate one.
It is only exact at tree level and explicitly broken by radiative corrections in the effective K\"ahler potential.
One-loop diagrams involving neutrino Yukawa couplings generate shift symmetry-violating terms such as $\delta K \simeq \big(\tilde h^\dagger \tilde h\big)/\left(16\pi^2\right) \left|N_1\right|^2$.
However, for Yukawa couplings of $ \order{0.1} $, these corrections are suppressed by a factor of $\mathcal{O}\!\left(10^{-4}\right)$, so that they are negligible for our purposes.

We also note that the symmetry governing the neutrino mass term in the superpotential is not affected by radiative corrections.
The Dirac mass term $M N_1 N_2$ is protected by the SUSY nonrenormalization theorem and thus radiatively stable.
A priori, we are thus allowed to assume zero (or arbitrarily small) Majorana masses, $\delta M_{1,2} = 0$.
There is no lower bound on $\delta M_{1,2}$ in consequence of radiative corrections to the superpotential.
As we will see later on, this parametric freedom in choosing the mass splitting between $M_1$ and $M_2$ will prove crucial to our analysis of resonant leptogenesis.

The suggested inflaton mass $ M \sim 10^{13} $ GeV is close to the scale preferred by the seesaw mechanism, with corresponding neutrino Yukawa couplings of $ \order{0.1} $.
This in turn implies a heavy neutrino decay rate of $ \order{10^{10}} $ GeV, which leads to a reheating temperature as large as $ 10^{14} $ GeV.
A reheating temperature of $ T_{R} \simgt 10^{9} $ GeV indicates thermal leptogenesis is possible \cite{Giudice:2003jh,hep-ph/0401240}.
The superfields $ N_{1,2} $ are thus responsible for inflation, neutrino masses and leptogenesis.

For inflaton field values greater than $ \order{10}\, M_P$, leptons and Higgs take super-Planckian masses, and the effective field theory description of inflation breaks down. This may be remedied by considering the continuous shift symmetry (in the K\"ahler potential) breaking to a discrete one in the superpotential, invariant under $ \phi \rightarrow \phi + 2\pi f $. 
The resultant inflationary potential is given in terms of sine functions, and the neutrino Yukawa couplings are periodic in the inflaton field. Hence the size of the lepton and Higgs masses are kept under control; if $ f \simlt 10 M_P $, the lepton and Higgs masses do not exceed $ M_P $.
Finally, we note that this version of chaotic sneutrino inflation is more consistent with Planck data than traditional chaotic inflation due to the smaller values of tensor to scalar modes.

\subsection{Mass and Yukawa matrices}
At leading order, the RH neutrino mass matrix contains only off-diagonal (Dirac) terms.
We assume that this mass structure is protected by an approximate  symmetry in the sector responsible for the (dynamical) generation of the RH neutrino masses.
However, it is in principle possible for physical processes at a lower energy scale $ \Lambda \ll M $ to produce also diagonal (Majorana) mass terms like $ \delta M_1 N_1^2 $, $ \delta M_2 N_2^2 $. 
These Majorana masses can be made explicitly real and positive by phase transformations. 
In Section~\ref{sec:deltaM}, we will discuss a possible origin of such terms from SUSY breaking in supergravity.
For now, we simply consider a single free parameter $ \delta M $, assuming $ \delta M_1 = \delta M_2 $ for convenience.
Note that successful inflation requires $ \delta M \simlt 10^{-2} M $ as an upper bound \cite{1601.00192}.
We will see that a lower bound is set by leptogenesis.
The RH neutrino mass matrix is thus given by 
\begin{equation}
	M_R = \pmatr{\delta M & M \\ M & \delta M},
\end{equation}
with eigenvalues $ M_1 = |M - \delta M| $, $ M_2 = |M + \delta M| $. Conversely, $ \delta M $ is equal to half the mass difference, i.\,e., $ \delta M = \tfrac{1}{2} (M_2 - M_1) $.

We do not make any assumptions about the nature or origin of the neutrino Yukawa couplings; but we may parametrise the Yukawa matrix $ h_{\alpha i} $ so as to incorporate the current experimental data on neutrino mass and mixing, as done in \cite{hep-ph/0103065}. With two RH neutrinos, there are nine free parameters at the high scale \cite{1601.00192}, giving seven observables at the low scale: three mixing angles $ \theta_{ij} $, two mass-squared differences $ \Delta m^2_{ij} $, one Dirac phase $ \delta_{\rm CP} $ and one Majorana phase $ \varphi $ (the other Majorana phase is zero). 
We must also consider two mass orderings: Normal Ordering (NO), where $ 0 = m_1 < m_2 < m_3 $, and Inverted Ordering (IO), where $ 0 = m_3 < m_1 < m_2 $.

The Yukawa matrix can then be specified by known quantities, with two excess degrees of freedom.
Specifically, in the diagonal RH neutrino basis, $ \tilde{h}_{\alpha i} \rightarrow h_{\alpha i} $, we write
\begin{equation}
	v_u\, h_{\alpha i} = 
	i\, U_{\alpha \gamma}^{\ast}\,
	\sqrt{m_{\gamma}}\,
	(R^\mathrm{T})_{\gamma i}\,
	\sqrt{M_{i}}
	,
\label{eq:halphai}
\end{equation}
where $ v_u = v \sin \beta$, $ v \approx 175 $ GeV is the electroweak Higgs VEV, 
$ m_{\gamma} $ and $ M_{i} $ are the light and heavy neutrino mass eigenvalues respectively, and 
$ U $ is the PMNS matrix. 
The rotation matrix $ R $ \cite{hep-ph/0103065} is given in terms of a complex free parameter $ \xi $ by
\begin{align}
	R = 
	\left\{
	\begin{array}{l}
		\pmatr{0 & \cos \xi & \sin \xi \\ 0 & -\sin \xi & \cos \xi} \quad \mathrm{[NO]} \\[3ex]
		\pmatr{\cos \xi & \sin \xi & 0 \\ - \sin \xi & \cos \xi & 0} \quad \mathrm{[IO]}
	\end{array}
	\right. .
\label{eq:rcasas}
\end{align}
This parametrisation is consistent with that in \cite{Antusch:2011nz}. We omit the factor $ \zeta = \pm 1 $ there which categorises the ``branches'' of the parametrisation, as the choice of $ \zeta $ has no effect on our results. This is demonstrated in Appendix \ref{sec:branches}.

Note further that since there are only two RH neutrinos, one left-handed (LH) neutrino is necessarily massless. By comparison with global fits \cite{1512.06856}, we fix the light masses to be
\begin{equation}
	m_{\gamma} = \mathrm{diag}(m_1,m_2,m_3) =
	\left\{
	\begin{array}{l}
		\mathrm{diag}(0,8.66,49.6) ~\mathrm{meV~~[NO]} \\[1.5ex]
		\mathrm{diag}(48.7,49.5,0) ~\mathrm{meV~~[IO]}
	\end{array}
	\right. .
\end{equation}

Relevant to leptogenesis is the quantity $ h^\dagger h $, which is invariant under lepton flavour basis changes. In the degenerate limit $ M_1 \rightarrow M_2 $, we have
\begin{equation}
	(h^\dagger h)_{ij} = \frac{M}{v_u^2}\, (R^\ast)_{i \gamma}\, m_{\gamma}\, R^\mathrm{T}_{\gamma j} \,,
\end{equation}
which is entirely independent of the PMNS matrix $ U $. In the single-flavour approximation (i.\,e., neglecting flavour effects), the $ CP $ asymmetries are given in terms of $ h^\dagger h $ only. 
We will find that even when heavy flavour effects are taken into account, the impact of varying the $ CP $ phases is minor. 
Phenomenologically, it can usually be subsumed into minor shifts in the unconstrained parameter $ \xi $, which is inaccessible at low energies.
Hence, this model alone can neither predict nor constrain $ \delta_{\rm CP} $ (weakly constrained by data) and $ \varphi $ (completely unknown).%
\footnote{\label{fn:phases}Models with two RH neutrinos can predict particular values of the $ CP $-violating phases if we assume one or two texture zeroes in $ h_{\alpha i} $. For two examples, see \cite{tsutomu}, based on Occam's razor, and \cite{fred}, which invokes a flavour symmetry with vacuum alignment.}
The main free parameters of the theory are thus $ \delta M $, $ \xi $ and $ \tan \beta $.

\section{Resonant leptogenesis}
\subsection{Analytical approximation}
\label{sec:approx}
We wish to construct a simple analytical result to get a sense of how resonant leptogenesis manifests in our model. If we assume the neutrinos follow Boltzmann distributions, we may use results already present in the literature on supersymmetric resonant leptogenesis, which we summarise here.

In principle, the final lepton asymmetry in our scenario is subject to two different types of flavour effects operating on the charged-lepton flavours $e$, $\mu$, and $\tau$: (i) light flavour effects induced by the charged-lepton Yukawa interactions involving the down-type Higgs doublet $H_d$, and (ii) heavy flavour effects induced by the neutrino Yukawa interactions involving the up-type Higgs doublet $H_u$. 
Provided that $\tan \beta$ takes a small value, the charged-lepton Yukawa interactions are, however, out of thermal equilibrium at high temperatures, corresponding to what is usually referred to as the ``unflavoured'' or ``single-flavour'' regime. In this regime, light flavour effects are negligible, whereas heavy flavour effects may still play an important role. More precisely, the condition of realising the ``unflavoured regime'' amounts to a lower bound on the heavy-neutrino mass~\cite{Abada:2006fw},
\begin{align}
	M \gg 5 \times 10^{11} \left(1 + \tan^2\beta \right) \,\textrm{GeV}.
\label{eq:tanbeta}
\end{align}
Recalling that $ M_{2,3} \sim 10^{13} $ GeV, this bound is obeyed provided $ \tan \beta \simlt 5 $.

As long as the condition in Eq.~\ref{eq:tanbeta} is satisfied, all light flavour effects are negligible. Let us, for now, also neglect all heavy flavour effects for simplicity.%
\footnote{We will devote more attention to heavy flavour effects in the next section, showing that these effects can become important.}
In this case, the final comoving $B-L$ number density is simply given by
\begin{equation}
	N^{B-L} = \sum_{i=1,2} \varepsilon_i\, \kappa_i \,,
\end{equation}
where $ \varepsilon_i $ is the CP asymmetry parameter and $ \kappa_i $ is the washout factor from inverse decays and scatterings evaluated at late time \cite{hep-ph/0401240,hep-ph/0605281}.
In a generic supersymmetric setup, there will be contributions of the form above coming from each of neutrino and sneutrino decays. The above expression also neglects phantom terms \cite{1003.5132}.

The unflavoured $ CP $ asymmetry parameters $ \varepsilon_i $ have been calculated in \cite{Covi:1996wh}, 
\begin{equation}
	\varepsilon_{i} = 
	\mathcal{H}_{i}\, \frac{M_i\Gamma_j}{M_i^2 - M_j^2} 
	\,, \quad 
	\mathcal{H}_{i} = \frac{\textrm{Im}\left\{\left[(h^\dagger h)_{ij} \right]^2 \right\}}
	{(h^\dagger h)_{ii}(h^\dagger h)_{jj}}.
\label{eq:Hi}
\end{equation}
where $ \Gamma_j $ is the total tree-level zero-temperature decay rate of heavy-neutrino species $N_j$. In SUSY, we have
\begin{equation}
	\Gamma_j = \frac{(h^\dagger h)_{jj}}{4\pi} M_j.
\end{equation}
At the SUSY level, the $CP$ asymmetries for neutrino and sneutrino decays are equal, $\varepsilon_i = \tilde{\varepsilon}_i$. In the following, we will assume that this relation is not disturbed by soft SUSY breaking effects.

The expression for $\varepsilon_i$ clearly diverges in the degenerate limit. This divergence is related to the heavy-neutrino self-energy diagram and needs to be regulated:
\begin{align}
	\frac{M_i\Gamma_j}{M_i^2 - M_j^2} \rightarrow 
	\frac{\left(M_i^2 - M_j^2\right)M_i\Gamma_j}{\left(M_i^2 - M_j^2\right)^2 + R_{ij}^2}.
\end{align}
The correct choice of the regulator $R$ is the subject of an ongoing debate in the literature \cite{Dev:2014laa,Dev:2014wsa,1406.4190,1404.4816,Kartavtsev:2015vto}.
For a comparison of results for $R$ in the literature, see Appendix~A of \cite{Dev:2014laa}. 
In the following, we will employ a possible form of the regulator, derived in the Kadanoff-Baym formalism from heavy-neutrino oscillations, which a large number of groups agrees upon:%
\begin{align}
	R_{ij} = 
	\left(M_i\Gamma_i+M_j\Gamma_j\right)
	\left[\frac{\det\left[\textrm{Re}\left\{h^\dagger h\right\}\right]}
	{\left(h^\dagger h\right)_{ii}\left(h^\dagger h\right)_{jj}}\right]^{1/2} .
\label{eq:regulator}
\end{align}
In summary, the asymmetry parameter related to heavy neutrino oscillations is given by
\begin{align}
	\varepsilon_{i}^{\rm osc} = 
	\frac{\textrm{Im}\left\{\left[(h^\dagger h)_{ij} \right]^2 \right\}}
	{(h^\dagger h)_{ii}(h^\dagger h)_{jj}}
	\,
	\frac{\left(M_i^2 - M_j^2\right)M_i\Gamma_j}{\left(M_i^2 - M_j^2\right)^2 + R_{ij}^2}.
\label{eq:epsiloni}
\end{align}
Note that due to the presence of the regulator, the asymmetry vanishes in the limit of degenerate (s)neutrino masses $ M_1 \rightarrow M_2 $. Moreover, for small mass splitting, $ \varepsilon_i^\mathrm{osc} $ is essentially linear in $ \delta M $.

It has been argued by one group (see e.\,g. \cite{Dev:2014laa}) that, in addition to the heavy-neutrino oscillations, $CP$ violation from heavy-neutrino mixing results in an asymmetry parameter of roughly the same order, $ \varepsilon_{i}^{\rm mix} $, for each heavy neutrino index $ i $. This has been disputed \cite{1406.4190} and is currently under discussion~\cite{Kartavtsev:2015vto}. We do not attempt to settle this question here, rather we will use the $CP$ asymmetry given in Eq.~\ref{eq:epsiloni}, with the caveat that there may be an additional factor of $ \sim 2 $. This will not significantly impact our analysis.

The washout factors $ \kappa_i $ for nearly degenerate neutrinos can be found in \cite{hep-ph/0603107}, which we quote here:
\begin{equation}
\begin{split}
\kappa_1 &  = \frac{2}{z_{12}\, K_{12}}  \\
\kappa_2 & = \frac{2}{z_{21}\, K_{21}}
\left[1+2 \ln\left(\frac{1+\delta}{1-\delta}\right)\right]^2
\exp\left[-\frac{3\pi}{8} K_1 \left(\frac{\delta}{1+\delta}\right)^{2.1}\right] ,
\end{split}
\end{equation}
where $ \delta = 2 \, \delta M / M $ and
\begin{equation}
	K_{ij} = K_i + K_j^{1-\delta} , \quad
	z_{ij} = z_B\!\left(K_i + K_j^{\left(1-\delta\right)^3}\right) , \quad
	z_B\!\left(K\right) \simeq 2 + 4 K^{0.13} e^{-2.5/K}.
\end{equation}
The decay parameters $ K_i $ are given by 
\begin{equation}
	K_i \equiv \frac{\Gamma_{N_i} (T=0)}{H(T=M_i)} = \frac{v_u^2 (h^\dagger h)_{ii}}{M_i m_*}, 
	\quad 
m_* \simeq (0.78 \times 10^{-3} \mathrm{~eV}) \sin^2 \beta.
\label{eq:Ki}
\end{equation}
In the limit $ \delta \rightarrow 0 $, this simplifies to
\begin{align}
	\kappa_1 = \kappa_2 = \kappa\left(K_1 + K_2\right) \,, \quad
	\kappa \left(K\right) = \frac{2}{z_B\left(K\right) K}.
\end{align}
This allows us to settle $ N^{B-L} $, but to compare this to observation, we need to convert this into a baryon density taking into account sphaleron processes. The present-day baryon-to-photon ratio, i.\,e., the present-day baryon asymmetry $ \eta_B $ is given by
\begin{equation}
	\eta_B = d^{\rm MSSM} \sum_i \varepsilon_i \kappa_i = d^{\rm MSSM} (\varepsilon_1 + \varepsilon_2) \, \kappa(K_1 + K_2),
\label{eq:etaB}
\end{equation}
where $ d^{\rm MSSM} \approx 0.89 \times 10^{-2} $ \cite{hep-ph/0406115}. 

\subsection{Full density matrix equations}
\label{sec:exact}
The above results depend on several simplifying assumptions, including the assumption that the flavour composition of the linear combinations of charged leptons $ \ell_1 $ and $ \ell_2 $, coupling respectively to $N_1$ and $N_2$, is irrelevant (i.\,e., no heavy flavour effects). It is further assumed that leptogenesis occurs at high enough temperatures that the coherence of propagating lepton states is not disturbed by charged-lepton Yukawa interactions (i.\,e., no light flavour effects). The latter relies on the assumption in Eq.~\ref{eq:tanbeta}, equivalent to $\tan\beta \simlt 5$, which we shall assume again.

However, we cannot safely neglect heavy flavour coherence effects in the degenerate limit. In the hierarchical limit, one may establish distinct phases of decay and washout in different directions of flavour space. Washout is typically then controlled by the flavoured decay factors $ K_{i \alpha} $. This has been studied extensively in \cite{1007.1641}. 

In our model, however, decay and washout happens simultaneously in all directions of flavour space, and we cannot assume a diagonal density matrix. Indeed, we expect coherence effects to affect the washout efficiency, which requires considering the full density matrix for the $ B-L $ asymmetry, $ N^{B-L}_{\alpha \beta} $, which accounts for all neutrino flavour effects.

The evolution equation for the $ B-L $ asymmetry matrix has been derived in \cite{1112.4528}. It may be written as
\begin{align}
	\frac{d}{dz} N_{\alpha\beta}^{B-L} = 
	\sum_{i=1,2} \left[
	\varepsilon_{\alpha\beta}^{(i)}\, D_i \left(N_{N_i} - N_{N_i}^{\rm eq}\right) -
	\frac{1}{2}\, W_i \left\{P_i^0,N_{B-L}\right\}_{\alpha\beta}
	\right] .
\label{eq:DME}
\end{align}
where $ z = M/T \simeq z_i = M_i/T $, $ \alpha $ and $ \beta $ are flavour indices and $ i $ indexes the right-handed (s)neutrinos. It is most convenient to study the evolution of this set of equations in the charged lepton flavour basis, as the Yukawa couplings in this basis correspond directly to the Casas-Ibarra parametrisation employed above.
The entries of $N_{\alpha\beta}^{B-L}$ then describe the physical flavour asymmetries and coherences, i.\,e.,
\begin{align}
	N_{\alpha\beta}^{B-L} = 
	\begin{pmatrix}
		N_{ee}^{B-L} & N_{e\mu}^{B-L} & N_{e\tau}^{B-L} \\
		N_{\mu e}^{B-L} & N_{\mu\mu}^{B-L} & N_{\mu\tau}^{B-L} \\
		N_{\tau e}^{B-L} & N_{\tau \mu}^{B-L} & N_{\tau\tau}^{B-L}
	\end{pmatrix} .
\end{align}
This asymmetry matrix is hermitian, so that it only contains six independent degrees of freedom. The entries on the diagonal of $N_{\alpha\beta}^{B-L}$ correspond to the three physical flavour asymmetries,
\begin{align}
	N_e^{B-L} \equiv N_{ee}^{B-L} , \quad
	N_\mu^{B-L} \equiv N_{\mu\mu}^{B-L} , \quad
	N_\tau^{B-L} \equiv N_{\tau\tau}^{B-L} ,
\end{align}
while the off-diagonal entries account for the coherences among the different flavour states,
\begin{align}
	N_{e\mu}^{B-L} = \left(N_{\mu e}^{B-L}\right)^* , \quad
	N_{e\tau}^{B-L} = \left(N_{\tau e}^{B-L}\right)^* , \quad
	N_{\mu\tau}^{B-L} = \left(N_{\tau \mu}^{B-L}\right)^* .
\end{align}
The total asymmetry is the trace of the asymmetry matrix,
$ \textrm{Tr}\!\left[N_{\alpha\beta}^{B-L}\right] = \textstyle{\sum_\alpha} N_\alpha^{B-L}$.

We now describe each term on the right-hand side in Eq.~\ref{eq:DME}.
The $CP$ asymmetry matrices $\varepsilon_{\alpha\beta}^{(i)}$ are given by
\begin{align}
	\varepsilon_{\alpha\beta}^{(i)} = 
	\mathcal{H}_{\alpha\beta}^{(i)} \,
	\frac{\left(M_i^2 - M_j^2\right)M_i\Gamma_j}
	{\left(M_i^2 - M_j^2\right)^2 + R_{ij}^2} ,
\end{align}
where the regulator $ R_{ij} $ is given,%
\footnote{As in Section \ref{sec:approx}, to remain conservative, we will work with this regulator only, which is agreed upon by all groups working on resonant leptogenesis.}
as in our analytical result, by Eq.~\ref{eq:regulator}, while $\mathcal{H}_{\alpha\beta}^{(i)}$ is 
\begin{align}
\thinmuskip=0mu
	\mathcal{H}_{\alpha\beta}^{(i)} = \frac{i}{2}\frac{
	h_{\alpha i}^{\vphantom{*}}h_{\beta j}^*\big(h^\dagger h\big)_{ji} -
	h_{\alpha j}^{\vphantom{*}}h_{\beta i}^*\big(h^\dagger h\big)_{ij} +
	\frac{M_i}{M_j} \left[
	h_{\alpha i}^{\vphantom{*}}h_{\beta j}^*\big(h^\dagger h\big)_{ij} -
	h_{\alpha j}^{\vphantom{*}}h_{\beta i}^*\big(h^\dagger h\big)_{ji}
	\right]}
	{\left(h^\dagger h\right)_{ii}\left(h^\dagger h\right)_{jj}} .
\end{align}
We note that $ \mathcal{H}_{\alpha \beta}^{(i)} $ is a generalization of the corresponding expression in the hierarchical approximation, $ \mathcal{H}_{i \alpha} $ (see e.\,g. \cite{1512.06739}), which in turn yields $ \mathcal{H}_i $ in Eq.~\ref{eq:Hi}. These expressions are related by $ \mathcal{H}_i = \sum_\alpha \mathcal{H}_{i \alpha} $ and $\mathcal{H}_{i \alpha} = \mathcal{H}_{\alpha \alpha}^{(i)} $, respectively.
Unlike $ \mathcal{H}_i $ in the last section, $ \mathcal{H}_{\alpha \beta}^{(i)} $ is implicitly dependent on PMNS parameters, in the off-diagonal elements $ \alpha \neq \beta $.

The decay terms $D_i$ are given by
\begin{align}
	D_i(z) = z\,K_i\,\frac{\mathcal{K}_1(z)}{\mathcal{K}_2(z)} \,,
\end{align}
where $ \mathcal{K}_{1,2} $ are modified Bessel functions of the second kind, of order $1,2$, and the decay parameters $ K_i $ are given in Eq.~\ref{eq:Ki}.
$N_{N_i}$ denotes the comoving number density of $N_i$ and obeys the Boltzmann equation
\begin{align}
	\frac{d}{dz} N_{N_i} = - D_i \left(N_{N_i} - N_{N_i}^{\rm eq}\right) , \quad
	N_{N_i}^{\rm eq} = \frac{z^2}{2} \mathcal{K}_2(z) ,
\end{align}
where $N_{N_i}^{\rm eq}$ is the comoving number density of a complete $N_i$ neutrino supermultiplet in thermal equilibrium. All comoving number densities are normalized such that $N_{N_i}^{\rm eq} \rightarrow 1$ in the ultra-relativistic limit $z\ll1$ (in the Boltzmann approximation).

The washout factors $W_i$ are given by
\begin{align}
	W_i(z) = \frac{z^3}{4} K_i \, \mathcal{K}_1(z).
\end{align}
The operators $P_i^0$ in the anticommutator in Eq.~\ref{eq:DME} are projection operators that project any given flavour state $ \ell_\alpha $ ($ \alpha = e,\mu,\tau $) onto the axes parallel to the linear combinations $\ell_i$. 
Given here at tree level, they are defined by the matrices $ P_i^0 = \ket{i}\bra{i} $, and may be written in compact matrix form in terms of tree-level amplitudes $ C_{\alpha i}^0 = \left<\alpha|i\right>$ as
\begin{align}
	\left(P_i^0\right)_{\alpha\beta} = C_{\alpha i}^0 C_{\beta i}^{0*} , \quad 
	C_{\alpha i}^0 = \frac{h_{\alpha i}}{\left(h^\dagger h\right)_{ii}^{1/2}},
\end{align}
The matrix elements of the anti-commutator $\left\{P_i^0,N_{B-L}\right\}$ thus take the form
\begin{align}
	\left\{P_i^0,N_{B-L}\right\}_{\alpha\beta} 
	= \frac{1}{\left(h^\dagger h\right)_{ii}}
	\sum_\gamma \left(h_{\alpha i}^{\vphantom{*}} h_{\gamma i}^{*} N_{\gamma\beta}^{B-L}
	+ N_{\alpha\gamma}^{B-L} h_{\gamma i}^{\vphantom{*}} h_{\beta i}^*\right) .
\end{align}

Finally, we establish the initial conditions. As we will argue shortly, we expect negligible contributions to the asymmetry from non-thermal inflaton decays, such that all asymmetry is produced by resonant thermal leptogenesis. This is equivalent to setting $ N^{B-L}_{\alpha \beta}(z\ll1) = 0 $. Furthermore, we assume thermal initial neutrino abundances, i.\,e., $ N_{N_i}(z\ll1) = N_{N_i}^\mathrm{eq}(z\ll1) $.
This is justified by noting that reheating after inflation is nothing other than the thermalisation of the energy content stored in the (sneutrino) inflaton field \cite{1601.00192}.

The final lepton asymmetry follows from evaluating the solution to Eq.~\eqref{eq:DME} at late times $z_f \gg 1$, such that
\begin{align}
	N_{\rm fin}^{B-L} 
	= N_e^{B-L}\left(z_f\right) + N_\mu^{B-L}\left(z_f\right) + N_\tau^{B-L}\left(z_f\right) .
\end{align}

\subsection{Vanishing initial asymmetry}
\label{sec:nonthermal}
Given a reheating temperature that is close to the mass of the sneutrinos, one may imagine that a certain proportion of the sneutrino decays occur non-thermally, an effect which would need to be captured beyond thermal leptogenesis.

It turns out that any non-thermal contributions are negligible, owing in part to the discrete shift symmetry.
As discussed in \cite{1601.00192}, the inflaton takes large field values during reheating, resulting in very heavy lepton and Higgs, such that perturbative inflaton decays like $ \phi \rightarrow \ell_i H $ are kinematically forbidden at this stage. 
Any non-thermal asymmetry must then be produced during the preheating phase. 

However, as we will justify in Section \ref{sec:deltaM} when considering SUSY breaking, the mechanism that produces non-zero $ \delta M $ (and thus non-zero $ CP $ asymmetry) is not active during this phase, so that lepton number $ L $ is not violated. We therefore conclude that any asymmetry produced non-thermally is negligible.

\section{Parameter space analysis}
\label{sec:numerics}

The primary free parameters under consideration here are:
\vspace{-3ex}
\begin{itemize}
\renewcommand{\itemsep}{-0.5ex}
	\item $ \delta M $: size of the mass degeneracy breaking for RH neutrinos,
	\item $ \xi $: complex argument of the matrix $ R(\xi) $ as defined in Eq.~\ref{eq:rcasas}, which parametrises the excess degrees of freedom in the neutrino Yukawa matrix.
\end{itemize}
\vspace{-3ex}
It is convenient to consider the dimensionless parameter $ \delta M / M $, such that a mass splitting of \ord(TeV) corresponds to $ \delta M/M \sim 10^{-10} $. 

In addition $ \tan \beta $ may also be varied, though we restrict ourselves to cases where $ \tan \beta \simlt 5 $, to observe the bound in Eq.~\ref{eq:tanbeta}. 
In this range, we find the total BAU $ \eta_B $ is essentially proportional to $ \sin^4 \beta $ for the analytical approximation (two powers each arise from $ \varepsilon_i $ and $ \kappa $). The full numerical solution shows $ \eta_B \propto \sin^5 \beta $. 
The origin of the extra factor $ \sin \beta $ is unclear.
Any extension of the model may place constraints on the allowed values of $ \tan \beta $. As an example, if SUSY breaking in the hidden sector is mediated to the visible sector only via gravitational interactions (such as in the mediation scheme of pure gravity mediation), achieving the correct Higgs mass typically requires a reasonably small $ \tan \beta $ \cite{Evans:2013dza}.
For consistency, all figures show results with $ \tan \beta = 5 $, giving $ \sin \beta \approx 0.98 $. 

Furthermore, we will find that the full solution to the density matrix equations contains a dependence on the (unknown or weakly constrained) $ CP $ phases of the PMNS matrix. However, this dependence is not strong; for any given $ \xi $, it may change the predicted $ \eta_B $ by an \ord(1) factor. More importantly, the excess of free parameters characterised by $ \xi $ implies the effect of varying the $ CP $ phases can typically be accounted for by a corresponding shift in $ \xi $. As this effect is comparatively small, we defer a more dedicated parameter space analysis, taking into account the effects of PMNS parameters, to a future work. For consistency, we choose
\begin{equation}
	\delta_{\rm CP} = 0, \qquad \varphi = 0.
\end{equation}
This choice, while not preferred by experiment, allows us to most easily compare analytical and numerical results.

Finally, while the neutrino mass-squared differences $ \Delta m^2_{21} $ and $ \Delta m^2_{31} $ are known, the absolute neutrino masses as appear in $ m_{\gamma} $ in Eq.~\ref{eq:halphai} depend on the mass ordering, i.\,e., the sign of $ \Delta m^2_{31} $. We consider both scenarios.

A comparison between the analytical approximation in Eq.~\ref{eq:etaB} and numerical solution to Eq.~\ref{eq:DME} reveals that they generally predict different signs of the total asymmetry $ \eta_B $. By examining the partial contributions from each flavour, $ N^{B-L}_\alpha $, $ \alpha = e, \mu, \tau $ in the numerical solutions, we find that the $ \tau $ flavour asymmetry agrees well with the approximation, including overall sign. However, when heavy flavour effects are switched on, the $ \mu $ flavour asymmetry goes from positive to negative.
As this is the dominant contribution to $ \eta_B $, the sign of $ \eta_B $ also switches. In this model we can always choose the sign of $ \eta_B $ by the freedom in determining $ \xi $. Specifically, a positive asymmetry is achieved when $ \mathrm{Re}[\xi] $ and $ \mathrm{Im}[\xi] $ have opposite sign.

\begin{figure}[ht]
	\centering
	\begin{subfigure}{0.50\textwidth}
	\centering\includegraphics[height=5.5cm]{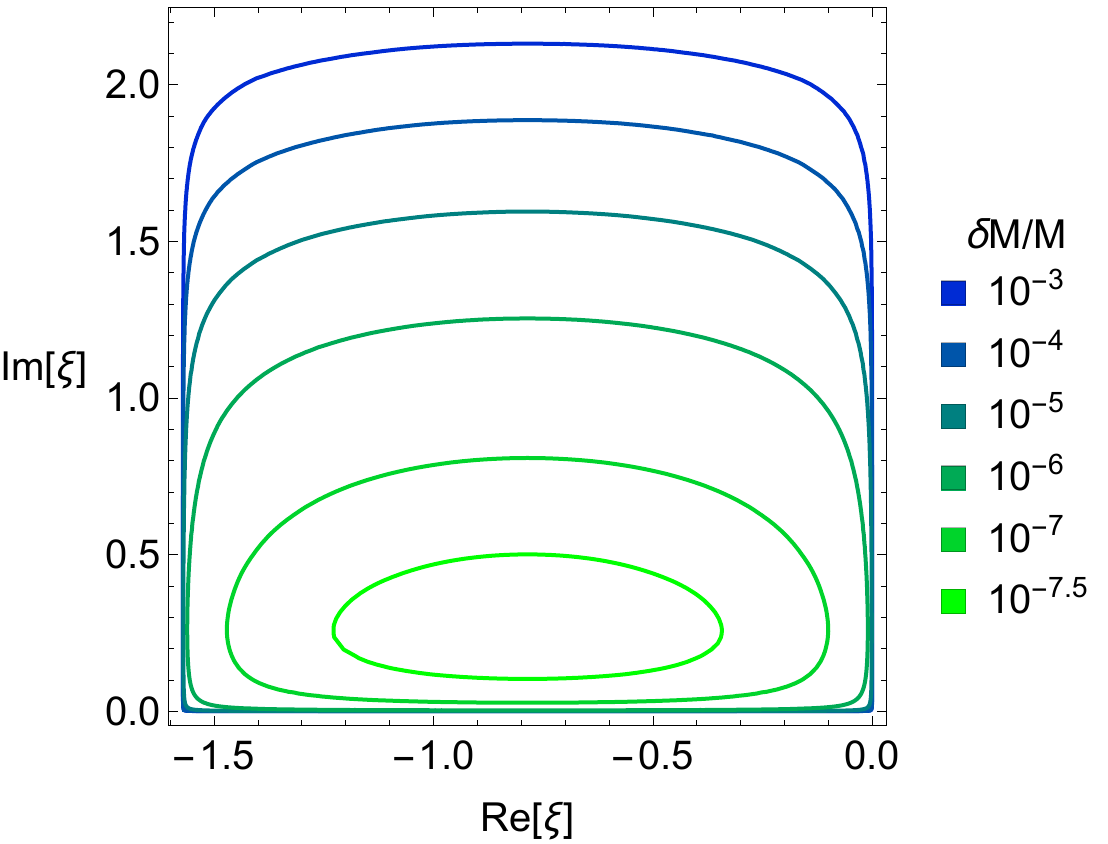}
	\end{subfigure}%
	\begin{subfigure}{0.50\textwidth}
	\centering\includegraphics[height=5.5cm]{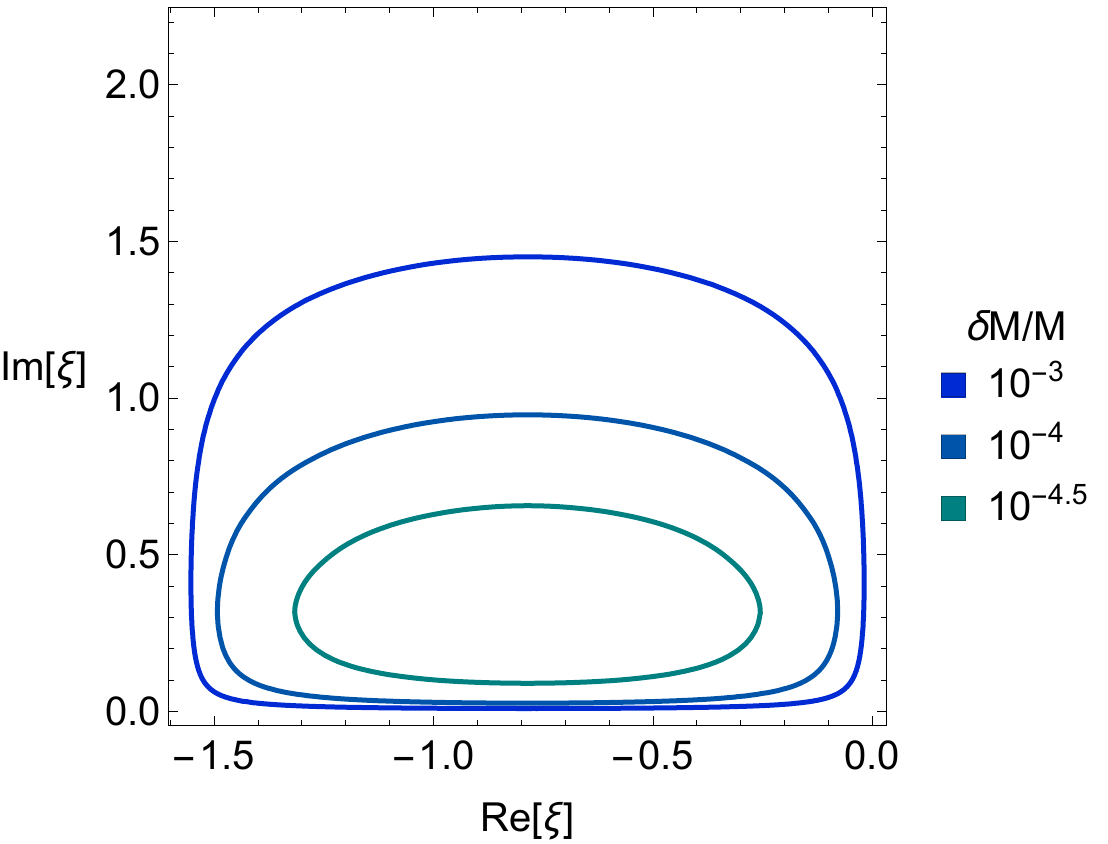}
	\end{subfigure}
	\caption{Contours where $ \eta_B = \eta_B^{\mathrm{obs}} $ for NO (left) and IO (right), as a result of solving the density matrix equation Eq.~\ref{eq:DME}. Note that, for $ \mathrm{Im}[\xi] > 0 $, achieving the correct sign of $ \eta_B $ requires $ \mathrm{Re}[\xi] < 0 $.}
	\label{fig:eta_vs_xi}
\end{figure}

Fig.~\ref{fig:eta_vs_xi} shows contours of $ \eta_B = \eta_B^\mathrm{obs} $ in terms of the complex parameter $ \xi $, for several values of $ \delta M $. We note in particular that the scale of $ \delta M $ is strongly dependent on the neutrino mass ordering; for NO, the correct asymmetry may be produced with $ \delta M/M \simgt 10^{-8} $, while IO requires $ \delta M/M \simgt 10^{-5} $. 
This will be significant when we discuss the origins of $ \delta M $ in Section~\ref{sec:deltaM}. 
The minimum $ \delta M $ giving the correct $ \eta_B $ corresponds to $ \delta M/M \approx 2.0 \times 10^{-8} $ (NO) and $ 1.5 \times 10^{-5} $ (IO), when $ \xi \approx \pm (\pi/4 - 0.3\,i) $.
For NO, assuming an inflaton mass $ M \sim 10^{13} $ GeV gives minimum $ \delta M \sim 200 $ TeV. Note also that the BAU is insensitive to the overall sign of $ \xi $. 

\begin{figure}[ht]
	\centering
	\includegraphics[height=5cm]{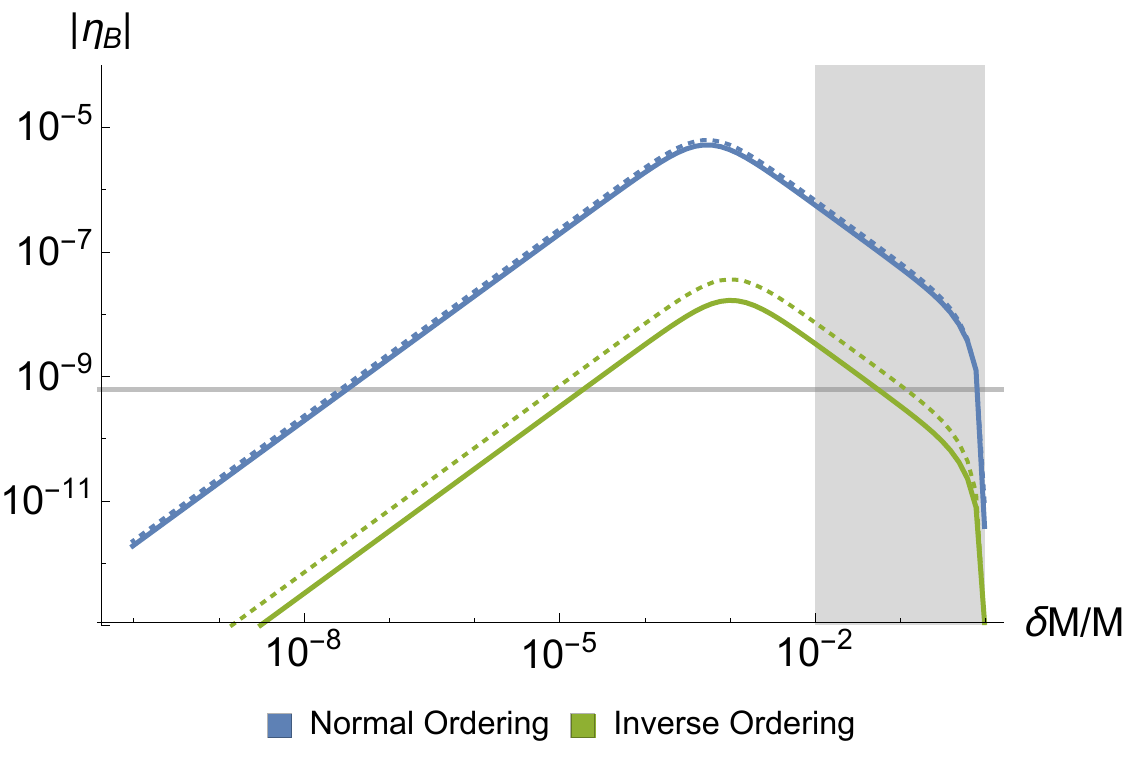}
	\caption{Baryon asymmetry plotted as a function of $ \delta M /M $, for $ \xi = -\pi/4 + 0.5\,i $. The shaded region at $ \delta M /M \simgt 10^{-2} $ is disallowed by inflation. Solid lines show exact (numerical) solutions to the density matrix equations in Section~\ref{sec:exact}, dotted lines show the approximation in Eq.~\ref{eq:etaB}.}
	\label{fig:eta_vs_dMM}
\end{figure}

Fig.~\ref{fig:eta_vs_dMM} shows the variation of the BAU with $ \delta M/M $ for fixed value $ \xi = -\pi/4 + 0.5\,i $. Solid lines show numerical solutions of Eq.~\ref{eq:DME}, while dotted lines plot Eq.~\ref{eq:etaB}. We note that for $ \delta M/M \simlt 10^{-4} $, $ \eta_B $ is essentially linear.
A maximal asymmetry is produced when $ 10^{-4} \simlt \delta M /M \simlt 10^{-3} $. For NO, this asymmetry may be over $ 10^4 $ times larger than the observed asymmetry, which would require another mechanism to wash out this excess. A third RH neutrino $ N_3 $ with mass $ M_3 \ll M_{1,2} $, though not necessary in this model, is not forbidden, and may provide the required washout.
Furthermore, the shaded region is disallowed by the inflation model, which requires a high degree of degeneracy between inflaton and stabiliser fields.

\begin{figure}[ht]
	\centering
	\begin{subfigure}{0.45\textwidth}
		\includegraphics[height=4cm]{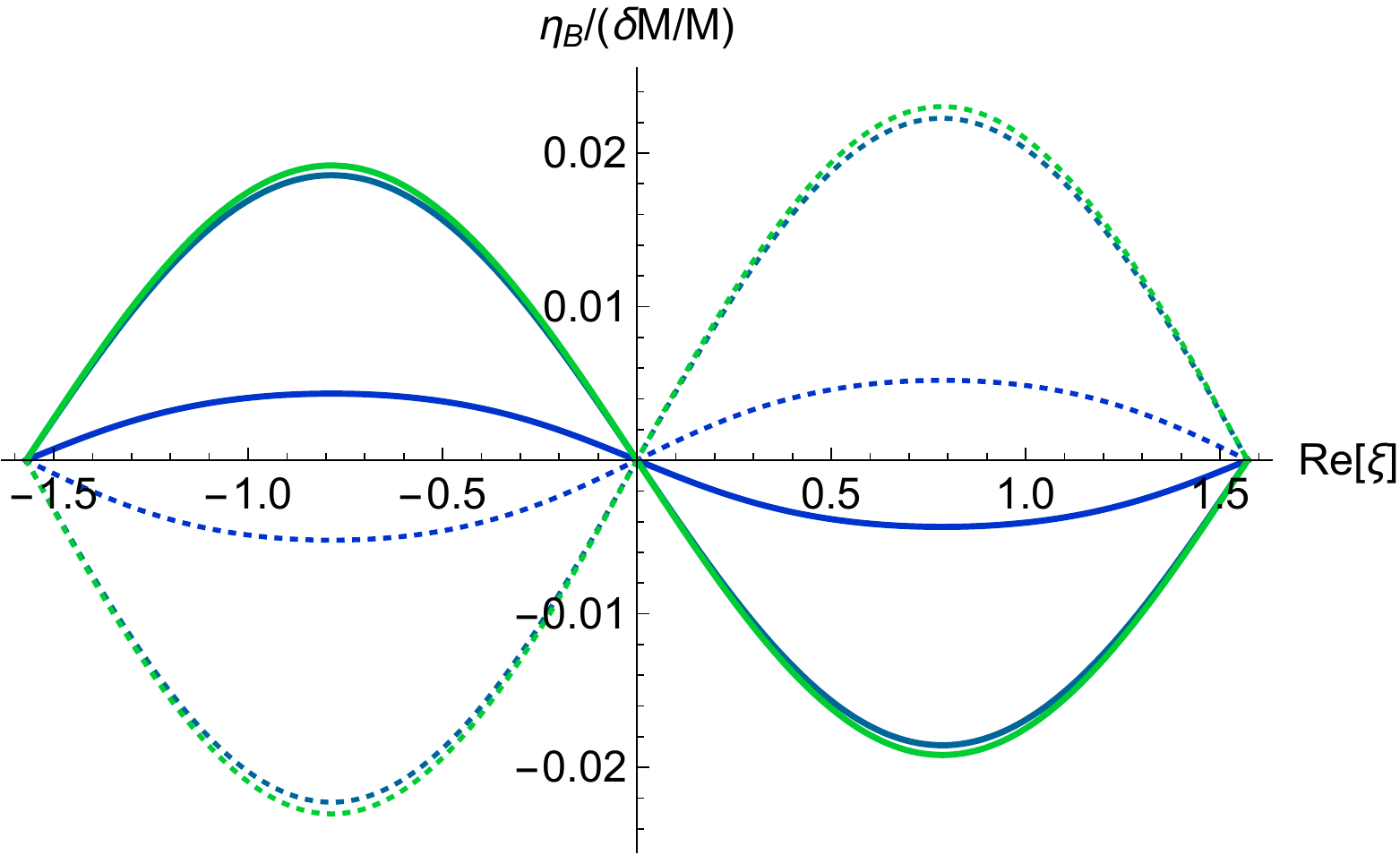}	
	\end{subfigure}%
	\begin{subfigure}{0.5\textwidth}
		\includegraphics[height=4cm]{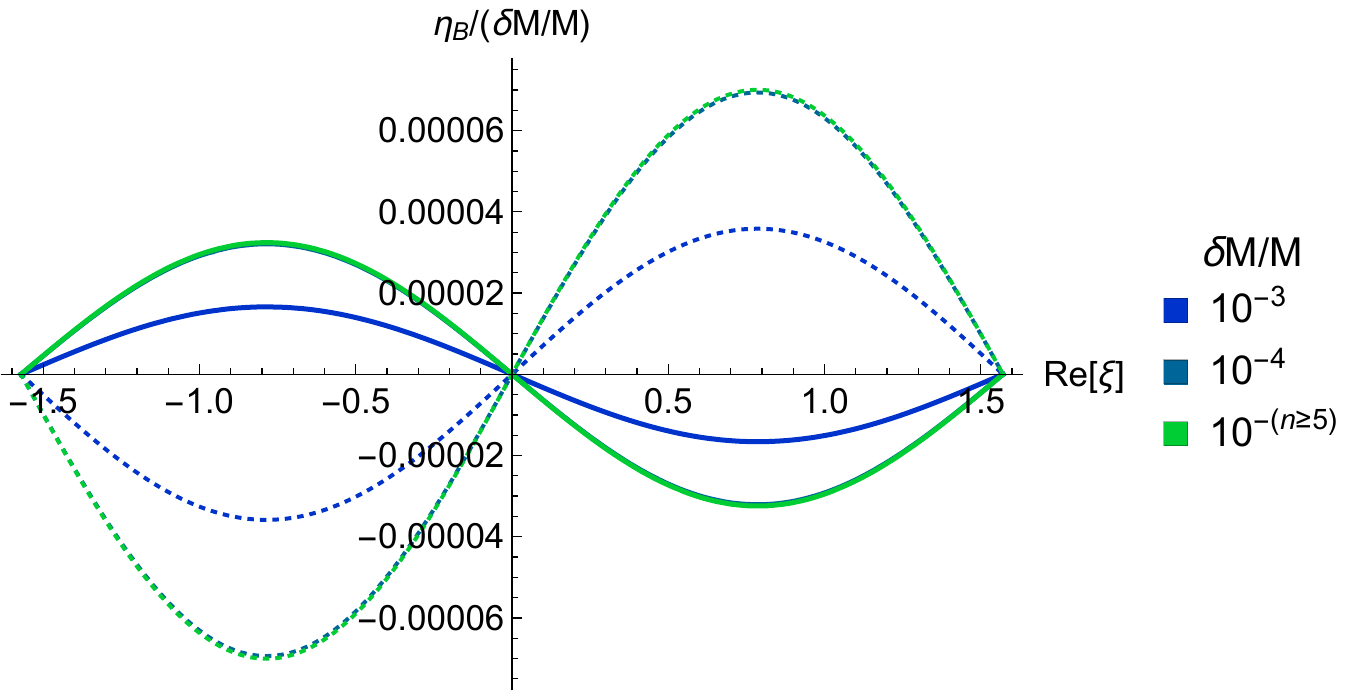}
	\end{subfigure}
	\caption{Baryon asymmetry (divided by $ \delta M/M $) against $ \mathrm{Re}[\xi] $, for NO (left) and IO (right). $ \mathrm{Im}[\xi] = 0.5 $. Solid lines show exact (numerical) solutions to the density matrix equations in Section~\ref{sec:exact}, dotted lines show the approximation in Eq.~\ref{eq:etaB}. Note that for $ \delta M/M \simlt 10^{-4} $, the scaled quantity $ \eta_B/(\delta M/M) $ is essentially constant in $ \delta M $ such that the plotted lines overlap.}
	\label{fig:eta_vs_rexi}
\end{figure}

Fig.~\ref{fig:eta_vs_rexi} shows the variation of $ \eta_B/(\delta M/M) $ with $ \mathrm{Re}[\xi] $, for $ \mathrm{Im}[\xi] = 0.5 $. It is convenient to consider a rescaled $ \eta_B $, as it removes the linear dependence on $ \delta M $ (for small $ \delta M $). As a consequence of the rescaling, several lines overlap in Fig.~\ref{fig:eta_vs_rexi}. 
A sinusoidal shape is immediately apparent; we find that $ \eta_B $ is proportional to $ \mp \sin(2\,\mathrm{Re}[\xi]) $. This stems from the fact that $\mathrm{Re}[(h^\dagger h)_{12}]$ is proportional to $\sin(2\,\mathrm{Re}[\xi])$. For $ \mathrm{Re}[\xi] \,\:\textrm{mod} \,\: \pi/2 = 0$, the real part of $(h^\dagger h)_{12}$ thus vanishes. $(h^\dagger h)_{12}$ is then purely imaginary, so that $[(h^\dagger h)_{12}]^2$ is in turn purely real. This results in vanishing $CP$ asymmetry parameters $\varepsilon_i$, which are proportional to $\mathrm{Im}\{[(h^\dagger h)_{12}]^2\}$. Hence a large asymmetry owing to large $ \delta M $ can be tuned to give the correct $ \eta_B $ by choosing precise values of $ \mathrm{Re}[\xi] $ close to $ 0 $ or $\pm \pi/2 $.

\begin{figure}[ht]
\centering
	\begin{subfigure}{0.5\textwidth}
		\includegraphics[height=4cm]{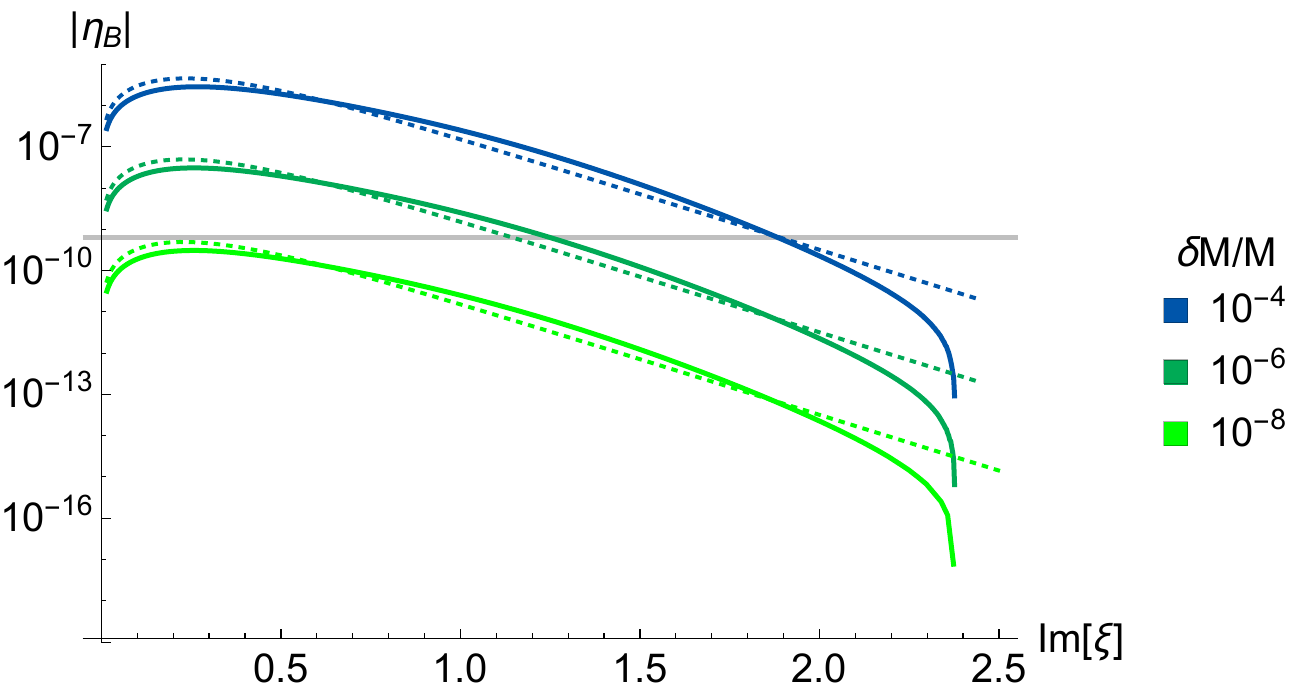}
	\end{subfigure}%
	\begin{subfigure}{0.5\textwidth}
		\includegraphics[height=4cm]{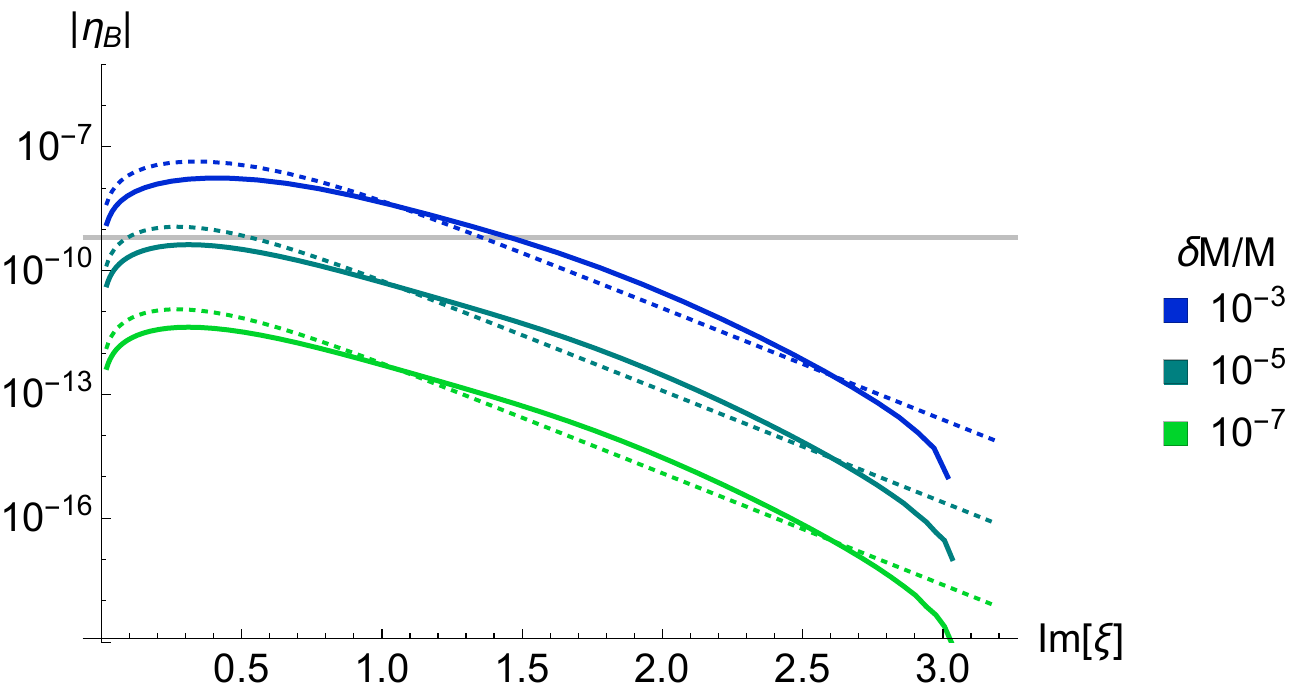}
	\end{subfigure}%
	\caption{Baryon asymmetry against $ \mathrm{Im}[\xi] $, for NO (left) and IO (right), for several choices of $ \delta M / M $. $ \mathrm{Re}[\xi] = - \pi/4 $. Solid lines show exact (numerical) solutions to the density matrix equations in Section~\ref{sec:exact}, dotted lines show the approximation in Eq.~\ref{eq:etaB}. Note the difference in plotted choices of $ \delta M/M $ between NO and IO; IO requires larger $ \delta M $ to attain the correct $ \eta_B $.}
	\label{fig:eta_vs_imxi}
\end{figure}

Fig.~\ref{fig:eta_vs_imxi} plots the asymmetry against $ \mathrm{Im}[\xi] $. Apart from the overall sign difference, here we see the largest discrepancy between the analytical approximation and the full numerical solution. 
The maxima at $ \mathrm{Im}[\xi] \approx 0.3 $ are consistent with Fig.~\ref{fig:eta_vs_xi}. 
Here we also see a severe drop-off develop at $ \mathrm{Im}[\xi] \approx 2.4 $ (NO) and $ \mathrm{Im}[\xi] \approx 3.1 $ (IO) in the numerical solutions. A possible explanation comes from noting that for large values of $ \mathrm{Im}[\xi] $, there are large Yukawa couplings $ h_{\alpha i} > 1 $. The two columns of the Yukawa matrix will be very similar, and we reproduce the low-scale neutrino data only due to cancellations between large terms. Barring a model that predicts this structure, such values are unpreferred. Furthermore, if $ h_{\alpha i} \gg 1 $, perturbativity is violated in the neutrino Yukawa couplings.

\section{Origin of \texorpdfstring{\boldmath{$\delta M$}}{delta M} from SUSY breaking}\label{sec:deltaM}

We have shown that the mass splitting, characterised by $ \delta M $, is central to understanding leptogenesis with two very massive and nearly degenerate RH neutrinos. It is thus important to understand its origins. We present here one compelling mechanism within the framework of supergravity, with important implications for the gravitino problem and the dark matter relic abundance. 

In supergravity, we may imagine that SUSY is broken in a hidden sector by the $ F $-term(s) of one or several new fields, and mediated to the visible sector by gravitational interactions. 
In the effective superpotential, these SUSY-breaking $F$-terms need to be balanced by a constant term of the form $m_0 M_P^2$, where $ m_0 $ is of the order of the gravitino mass $m_{3/2}$. This parameter needs to be tuned, so as to achieve a vanishing cosmological constant in the SUSY-breaking vacuum. 

The constant term in the superpotential leads to both a Majorana mass term $ m_0 N_1^2 $ for neutrinos and off-diagonal entries in the RH sneutrino mass matrix (where the diagonal elements are populated by terms of $ \order{M^2} $).

The simplest O'Raifeartaigh model of SUSY breaking via a nonzero $F$-term is the Polonyi model~\cite{Polonyi}. Its superpotential simply consists of two terms: a SUSY-breaking $F$-term, and a constant, which is required to tune the cosmological constant to zero. In Appendix \ref{sec:polonyi} we discuss the Polonyi model in more detail, and some important physical considerations when attempting to embed it into realistic particle physics models.
For the argument presented below, these details are largely irrelevant, although we will explicitly assume that $ m_0 = m_{3/2} $. How this equality may arise is shown in Appendix \ref{sec:polonyi}.

With the addition of $ m_{3/2} M_P^2 $, the relevant terms%
\footnote{We omit the quartic term in $ K $, and the neutrino Yukawa couplings $ \tilde{h}_{\alpha i} H^u L_\alpha N_i $ in $ W $.}
in the K\"ahler and superpotential (Eqs.~\ref{eq:Kinf}--\ref{eq:Winf}) may be written as
\begin{align}
	K &= |N_1|^2 + |N_2|^2 + \left( \frac{1}{2} N_1^2 + \mathrm{h.c.} \right) \,, \\
	W &= M N_1 N_2 + m_{3/2} M_P^2.
\end{align}
We bring the K\"ahler potential into canonical form by a K\"ahler transformation, giving
\begin{align}
	K &= |N_1|^2 + |N_2|^2, \label{eq:Knew}\\
	W &= e^{N_1^2 /2M_P^2} (M N_1 N_2 + m_{3/2} M_P^2),\label{eq:Wnew}
\end{align}
where the the K\"ahler metric is simply the unit matrix. 
The scalar potential $ V_F $ (shown in Eq.~\ref{eq:vsugra}) yields the following bilinear terms:
\begin{equation}
	V_F \supset M^2 |N_1|^2 + M^2 |N_2|^2 
	- m_{3/2} M (N_1 - N_1^\ast) (N_2 - N_2^\ast) + \order{m_{3/2}^2}
\label{eq:vfbilinears}
\end{equation}
The $ F $-terms for $ N_{1,2} $ give us (among other terms) the diagonal entries of the sneutrino mass matrix, $ M^2 |N_{1,2}|^2 $.
In addition, there are several bilinear ($ B $-) terms proportional to powers of $ m_{3/2} $ which couples $ N_1, N_1^\ast $ to $ N_2, N_2^\ast $, giving rise to mass splittings.
To demonstrate this rigorously, we split the complex fields into scalar and pseudoscalar components, $ N_i \rightarrow 1/\sqrt{2}\, (s_i + i p_i) $.
We construct the $ 4 \times 4 $ real mass matrix, to $ \order{m_{3/2}} $,
\begin{equation}
	V_F \supset \pmatr{s_1 \\ s_2 \\ p_1 \\ p_2}^\mathrm{T}
	\pmatr{
		M^2 & 0 & 0 & 0 \\
		0 & M^2 & 0 & 0 \\
		0 & 0 & M^2 & 2m_{3/2}M \\
		0 & 0 & 2m_{3/2}M & M^2
	}
	\pmatr{s_1 \\ s_2 \\ p_1 \\ p_2},
\label{eq:scalarmassmatrix}
\end{equation}
which has eigenvalues $ M^2, M^2, M(M \pm 2m_{3/2}) $.
Specifically, the mass degeneracy of the scalar components $ s_1, s_2 $ is broken only at $ \order{m_{3/2}^2} $, while the pseudoscalar components have eigenvalues $ M (M \pm 2 m_{3/2}) $. 
This is an immediate consequence of the special form of the K\"ahler potential with a shift symmetry in $ \mathrm{Im}[N_1] $.

Note that the above discussion only applies to sneutrinos; $ B $-terms cannot give fermions masses.
They arise directly from the superpotential in Eq.~\ref{eq:Wnew}.
Recall that fermion mass terms are derived from a superpotential $ W $ by
\begin{equation}
	\mathcal{L} \supset - \frac{1}{2} W_{ij} \psi^i \psi^j + \mathrm{h.c.} , \quad 
	W_{ij} = \left. \frac{\partial^2 W}{\partial \Phi^i \partial \Phi^j} \right|_{\Phi^i \rightarrow \phi^i}.
\end{equation}
This results in a fermion mass matrix $ m_{ij} = W_{ij} $, with
\begin{equation}
	m_{ij} = \pmatr{m_{3/2} & M \\ M & 0}.
\label{eq:massmatrix}
\end{equation}
The eigenvalues of $m^\dagger m$ are
\begin{equation}
	M^2 + \frac{1}{2}\,m_{3/2}^2 \pm m_{3/2} M\sqrt{1 + \left(\frac{m_{3/2}}{2\,M}\right)^2}
	\approx M (M \pm m_{3/2}).
\end{equation}

In summary, the RH neutrino and sneutrino mass squared eigenvalues, to $ \order{m_{3/2}} $, are

\begin{tabular}{ll}
\quad	$ M(M \pm m_{3/2}) $ & (neutrino),\\
\quad	$ M^2 $ & (sneutrino scalar component),\\
\quad	$ M(M \pm 2 m_{3/2}) $ & (sneutrino pseudoscalar component).
\end{tabular}

We see that $ CP $ asymmetries $ \varepsilon_i $ arise from neutrino and (pseudoscalar) sneutrino decays.%
\footnote{%
	A simpler mass structure can be achieved if we consider pure gravitational SUSY breaking \cite{Izawa:2010ym}. Here, the mass eigenvalues for both neutrinos and sneutrinos are equal, arising from a mass matrix like in Eq.~\ref{eq:massmatrix}. The analysis in Section \ref{sec:numerics} follows directly.
}
The physical conclusion we may draw is that in supergravity, where the SUSY breaking scale is associated with the gravitino mass $ m_{3/2} $, we naturally expect a splitting between the (s)neutrino masses at $ \order{m_{3/2}} $. We showed in Section \ref{sec:numerics} that $ \delta M / M $ can be $ \order{10^{-8}} $ for natural values of the complex phase $ \xi $, producing the correct BAU.
This assumes NO, which is preferred over IO by global fits. For IO, the corresponding scale is $ \order{10^3} $ larger.

The analysis in Section \ref{sec:numerics} assumes equal contribution from fermion and scalar neutrino decays, while SUSY breaking in supergravity leads to non-uniform mass splittings between neutrinos and sneutrinos. 
Nevertheless, we expect approximately equal contributions. The scalar sector has half as many degrees of freedom contributing to the asymmetry, but a mass splitting that is twice as large as for fermions. Recalling that $ \eta_B \propto \delta M $ for small $ \delta M $, the factors of two cancel.
Assuming an inflaton/sneutrino mass of $ M \sim 2 \times 10^{13} $ GeV, a mass splitting of $ \delta M /M \approx 2\times 10^{-8} $ corresponds to $ m_{3/2} \approx 2 \, \delta M \simgt 800 $ TeV. This bound may be lowered by a factor of (approximately) two from neutrino flavour mixing effects (see discussion in Section \ref{sec:approx}), and possibly by an additional \ord(1) factor from particular choices of $ CP $ phases in the PMNS matrix.

In short, producing the baryon asymmetry of the universe at the correct scale implies a gravitino mass of \ord(100-1000) TeV! 
This has several important consequences.
We begin by noting that such a heavy gravitino is welcome for explaining the observed Higgs boson mass of $ 125 $ GeV \cite{Ibe:2011aa}.

As noted in \cite{1601.00192}, a gravitino mass at (or above) this scale is welcome with regards to the gravitino problem.
Typically, a high reheating temperature such as predicted by this model, $T_{R} \sim10^{14} $ GeV, leads to copious production of gravitinos, which, once gravitinos decay into matter, will spoil the precise predictions from Big Bang nucleosynthesis (BBN).
However, even the lightest gravitino allowed by this model is heavy enough that it decays into radiation before the BBN era.

The high reheating temperature and consequently large gravitino abundance nevertheless leads to an overproduction of the lightest supersymmetric particle (LSP). Assuming $ R $-parity, this leads to a dark matter (DM) relic abundance that is too large. To solve this problem, we must assume some small degree of $ R $-parity violation.

The order of magnitude estimate for $ m_{3/2} $ can be altered or improved by additions to this minimal model.
For instance, a specific model of SUSY breaking can give a different correspondence between $ m_0 $, $ m_{3/2} $ and $ \delta M $.
Furthermore, while the numerical results presented in Section \ref{sec:numerics} are precise, they are given in terms of a complex phase $ \xi $ which, for certain values of $ \xi $, can produce the correct asymmetry even when $ \delta M $ is large. An extended model that further constrains the neutrino Yukawa matrix would add predictivity. This could be achieved by considering a flavour symmetry (see also footnote \ref{fn:phases}).

Finally, a comment on the vanishing non-thermal lepton asymmetry as discussed in Section \ref{sec:nonthermal}.
Let us assume that the constant term in the superpotential is generated in consequence of dynamical $R$ symmetry breaking at a scale $\Lambda_R$. We then naively expect that $m_{3/2} M_P^2 \sim \Lambda_R^3$. As long as $m_{3/2}$ does not exceed values of $\mathcal{O}\!\left(1000\right)\,\textrm{TeV}$, the dynamical scale $\Lambda_R$ always remains smaller than the reheating temperature, $\Lambda_R \lesssim T_R \sim 10^{14}\,\textrm{GeV}$. $R$ symmetry is therefore unbroken during reheating. This implies that the (s)neutrino mass spectrum is not yet split at this stage, so that lepton number $L$ is still a good quantum number at this time. Nonthermal processes during (p)reheating are therefore not capable of generating any lepton asymmetry, which justifies our assumption of setting the initial asymmetry in our analysis of resonant thermal leptogenesis to zero.

This argument is contingent on a rather precise hierarchy in $ \Lambda_R $ and $ T_R $, corresponding to a narrow range for $ m_{3/2} $ of $ \mathcal{O}\!\left( 100 - 1000 \right) $ TeV. We note that the upper bound, which is proportional to $ T_R^3 $, carries a large uncertainty, as reheating is a non-perturbative process in this model (see discussion in \cite{1601.00192}). It is possible to have $ T_R \sim 10^{15} $ GeV, corresponding to $ m_{3/2} < \mathcal{O}\!\left( 10^5 \right) $ TeV. 
However, the case $ m_{3/2} > \mathcal{O}\!\left( 10^5 \right) $ TeV may allow a significant non-thermal contribution to the total asymmetry, in the absence of a mechanism which fixes the scale of $ R $ symmetry breaking.
As the pre-existing asymmetry is produced from the same interactions that give the thermal asymmetry, we naively expect them to have the same sign, such that the calculated asymmetry $ \eta_B $ in our earlier analysis amounts to a lower bound on the total asymmetry. 

\section{Conclusion}
\label{sec:conclusion}
In this paper we have explored resonant leptogenesis in a supersymmetric model with two heavy right-handed neutrinos $ N_{1,2} $, where their scalar parts act as the inflaton and stabiliser fields in a viable implementation of chaotic inflation. Successful inflation is achieved with a superpotential $ W = M N_1 N_2 $ with $ M 
\sim 10^{13} $ GeV, which implies degenerate right-handed neutrino masses. However, leptogenesis from (s)neutrino decays requires this degeneracy to be broken by some small amount $ \delta M \ll M $, which in turn controls the resonant enhancement.

With two RH neutrinos, the Yukawa matrix $ h_{\alpha i} $ is tightly constrained by experimental data on neutrino mixing angles and mass-squared differences. In the Casas-Ibarra parametrisation, the excess degrees of freedom in $ h_{\alpha i} $ are accounted for by a single complex phase $ \xi $. We have examined the dependence of the BAU on both $ 
\delta M $ and $ \xi $.

An analytical expression for the $ B-L $ asymmetry $ N^{B-L} $ given two nearly-degenerate neutrinos was derived in the Boltzmann approximation from known results (Eq.~\ref{eq:etaB}). However, this is unreliable for very small mass splittings, as we expect heavy neutrino flavour effects to play a significant role.
Against this benchmark, we numerically solved the exact evolution equation for the $ B-L $ asymmetry matrix $ N^{B-L}_{\alpha \beta} $ (Eq.~\ref{eq:DME}). 
To the authors' knowledge, this is the first time the calculation has been performed for two heavy,
nearly-degenerate neutrinos.

A comparison of analytical and density matrix solutions shows that there is an \ord(1) discrepancy in many regions of parameter space. In such cases, it may be acceptable to consider the simple analytical approximation. However, it completely fails to capture the correct physics in the case where the charged leptons $\ell_1$ and $\ell_2$ (that is, the linear combinations coupling to $N_1$ and $N_2$, respectively) are closely aligned in flavour space (corresponding to large $ \mathrm{Im}[\xi] $, see Fig.~\ref{fig:eta_vs_imxi}), and a density matrix approach must be used.

The correct BAU may be produced by very small mass splitting, where $ \delta M / M \sim 10^{-8} $ (NO) or $ 10^{-5} $ (IO).
In fact, small $ \delta M $ appears to be preferred, as this corresponds to generally small Yukawa couplings and no peculiar alignment in flavour space.
Small $ \delta M $ may be explained by considering SUSY breaking in supergravity, which comes with an additional term in the superpotential like $ m_{3/2} M_P^2 $. We have shown how this leads to a small Majorana mass for the fermionic RH neutrinos and produces a mass splitting between the pseudoscalar components of the sneutrinos. 

If $ \delta M / M \sim 10^{-8} $, this implies a gravitino mass of \ord(100-1000) TeV. This has 
important consequences for collider physics and cosmology. In particular, such a large gravitino mass implies that squarks and sleptons will not be discovered at the LHC, and also that the gravitino cosmological problem  is resolved.
We emphasise that this is the first paper which connects the baryon asymmetry of the universe to the SUSY breaking scale.

This model could be extended to include a third RH neutrino $ N_3 $ with a mass $ M_3 \ll M_{1,2} $. While thermal asymmetry from $ N_3 $ decays would be negligible, we may imagine a resonant enhancement of the $ CP $ asymmetry from $ N_{1,2} $ decays (if, say, $ \delta M / M \sim 10^{-4} $) resulting in too large a $B-L$ asymmetry by several orders of magnitude. This would need to be washed out by inverse decays of the third neutrino, and may result in bounds on the lightest RH neutrino mass. It would be interesting to consider this scenario in the future.

\subsection*{Acknowledgements}

The authors wish to thank Bhupal~Dev, Mathias~Garny, and Rasmus~Lundkvist for valuable discussions and comments. 
S.\,F.\,K. acknowledges the STFC Consolidated Grant ST/L000296/1 and the European Union's Horizon 2020 Research and Innovation programme under Marie Sk\l{}odowska-Curie grant agreements 
Elusives ITN No.\ 674896 and InvisiblesPlus RISE No.\ 690575.
This work has been supported in part by Grant-in-Aid for Scientific Research from the Ministry of Education, Culture, Sports, Science, and Technology (MEXT), Japan, Kakenhi No.\ 26104009 (T.\,T.\,Y.); Grants-in-Aid No.\ 26287039 and No.\ 16H02176 (T.\,T.\,Y.); and by the World Premier International Research Center Initiative (WPI), MEXT, Japan (T.\,T.\,Y.).
F.\,B.\ thanks Prof.\ Tsutomu T.\ Yanagida and the IPMU for their hospitality during his stay in Tokyo. The visit was funded by grant agreement InvisiblesPlus RISE No.\ 690575.

\appendix

\section{The Polonyi model of SUSY breaking}
\label{sec:polonyi}
In this Appendix we outline a minimal model of SUSY breaking in supergravity, based on the Polonyi model \cite{Polonyi}, and justify the assumption that $ m_0 = m_{3/2} $. Cosmological aspects are discussed.

We begin by noting that the $ F $-term part of the scalar potential in SUGRA is given by
\begin{equation}
	V_F = F^i \mathcal{K}_i^{\phantom{i}{\bar{\jmath}}} F_{\bar{\jmath}}^{\ast} - 3\, e^{K/M_P^2} \frac{|W|^2}{M_P^2}
\label{eq:vsugra}
\end{equation}
where $\mathcal{K}_i^{\phantom{i}{\bar{\jmath}}} $ is the K\"ahler metric and $ F_i $ is the generalised $ F $-term
\begin{equation}
	F^i = - \mathcal{K}^i_{\phantom{i}{\bar{\jmath}}}\: e^{K/2M_P^2} (D_j W)^\ast, \quad 
	\mathcal{K}^i_{\phantom{i}{\bar{\jmath}}} = (\mathcal{K}^{-1})^i_{\phantom{i}{\bar{\jmath}}}, \quad
	D_i W = \frac{\partial W}{\partial \phi^i} + \frac{W}{M_P^2} \frac{\partial K}{\partial \phi^i}.
\end{equation}
The Polonyi model assumes a single new chiral superfield $ X $. In its original form, the K\"ahler and superpotentials are given by
\begin{align}
	K &= |X|^2, \label{eq:polonyiK} \\
	W &= \mu^2 X + m_0\, M_P^2, \label{eq:polonyiW}
\end{align}
where $ \mu $ is the scale of SUSY breaking, and $ m_0 $ is an order parameter for $ R $-symmetry breaking. We emphasise that the second term in $ W $ breaks $ R $ but not SUSY; indeed, its dynamical origin may be completely different from SUSY breaking. In our universe, however, the two scales are linked by the requirement of a vanishing cosmological constant, $ \braket{V_F} = 0 $, which is required to recover the Minkowski vacuum that we observe.

SUSY is broken by the VEV of the auxiliary $ F $-term component of $ X $, i.\,e., $ \braket{F_X} = \mu^2 \neq 0 $.
Meanwhile, $ X $ acquires a VEV $ \braket{X} = x\,M_P $, where $ x $ is a dimensionless constant, expected to naturally be \ord(1).
The parameter $ m_0 $ is related to the gravitino mass $ m_{3/2} $ by an \ord(1) factor in terms of $ x $. Specifically, the gravitino mass is identified as follows
\begin{equation}
	m_{3/2} = e^{\left<K\right>/2/M_P^2}\,\frac{\left|W\right|}{M_P^2}
	= e^{\left|x\right|^2/2}\left|m_0 + x\, \frac{\mu^2}{M_P}\right|
\end{equation}
Imposing the vanishing cosmological constant condition, $ \braket{V_F} = 0 $, one finds in the original Polonyi model that $x = \sqrt{3}-1$ and $\mu^2 = \left(2 + \sqrt{3}\right) m_0\, M_P$. This yields
\begin{equation}
	m_{3/2} = e^{2-\sqrt{3}} \left(2+\sqrt{3}\right)\left|m_0\right| \approx 4.9\left|m_0\right|\,.
\end{equation}
This minimal realisation of SUSY breaking suffers from the ``cosmological Polonyi problem'' 
\cite{Coughlan:1983ci},
where a large VEV $ \braket{X} \sim M_P $ results in too much energy being stored in the Polonyi field oscillations after the end of inflation. 
This problem can be avoided in more realistic models which feature nonrenormalisable higher-order terms in $ X $ in the effective K\"ahler potential, such as $ \left|X\right|^4/M_*^2 $~\cite{Izawa:1996pk} or $ \left|X\right|^2\left|\Phi\right|^2/M_*^2 $~\cite{Linde:1996cx} for some cut-off scale $M_*$. For appropriate coefficients, the scalar potential in such models is minimised at $ \braket{X} = 0 $.

Eqs.~\ref{eq:polonyiK}--\ref{eq:polonyiW} then emerge as a low-energy effective theory of a more complete theory at higher energies. If $ x = 0 $, we have $ m_0 = m_{3/2} $ exactly. 
Although we have not specified an exact form of the extended Polonyi model, it provides a justification for the assumption made in the discussion in Section \ref{sec:deltaM}.

For consistency we must check that the inclusion of a new chiral superfield $ X $ does not lead to mixing between it and the neutrino superfields. 
We consider the ``$ N_i + X $'' theory, where the K\"ahler potential is given by the sum of K\"ahler terms as defined Eqs.~\ref{eq:Knew} and \ref{eq:polonyiK}, while the superpotential is the sum of terms in Eqs.~\ref{eq:Wnew} and \ref{eq:polonyiW}.
Calculating the scalar potential Eq.~\ref{eq:vsugra} in the $ N_i + X $ theory, we arrive at a $ 6\times6 $ mass matrix analogous to that in Eq.~\ref{eq:scalarmassmatrix}. We define $ X = s_x + i p_x $. 
For canonical K\"ahler potential and setting $ \braket{X} $ to zero, we find at $ \order{m_{0}^2} $,
\begin{equation}
	V_F \supset \pmatr{s_1 \\ s_2 \\ p_1 \\ p_2 \\ s_x \\ p_x}^\mathrm{T}
	\pmatr{
		M^2 + 4m_0^2 & 0 & 0 & 0 & 0 & 0\\
		0 & M^2 + m_0^2 & 0 & 0 & 0 & 0\\
		0 & 0 & M^2 & 2m_{0}M & 0 & 0 \\
		0 & 0 & 2m_{0}M & M^2 + m_0^2 & 0 & 0 \\
		0 & 0 & 0 & 0 & -2m_0^2 & 0\\
		0 & 0 & 0 & 0 & 0 & -2m_0^2
	}
	\pmatr{s_1 \\ s_2 \\ p_1 \\ p_2 \\ s_x \\ p_x}.
\end{equation}
The absence of off-diagonal entries coupling neutrinos to $ X $ shows there is no mixing.

\section{Branches of the Casas-Ibarra parametrisation}
\label{sec:branches}

In this appendix, we are going to expand in a bit more detail on the Casas-Ibarra parametrisation of the Yukawa couplings in Eq.~\ref{eq:halphai}. In particular, we will argue that our analysis in the main text actually covers the entire relevant parameter space, although we decide to focus on only one of two possible branches in the Casas-Ibarra parametrisation.

For both NH and IH, there are in fact two possible choices for the rotation matrix $R$~\cite{hep-ph/0103065},
\begin{align}
	R^{(\zeta)}\left(\xi\right) = 
	\left\{
	\begin{array}{l}
		\pmatr{0 & +\cos \xi & \zeta\,\sin \xi \\ 0 & -\sin \xi & \zeta\,\cos \xi}
		\quad \mathrm{[NO]} \\[3ex]
		\pmatr{+\cos \xi & \zeta\,\sin \xi & 0 \\ - \sin \xi & \zeta\,\cos \xi & 0}
		\quad \mathrm{[IO]}
	\end{array}
	\right. \,, \quad \zeta = \pm 1 \,,
\end{align}
where $\zeta = \pm1$ defines a ``positive branch'' and a ``negative branch'' in Eq.~\ref{eq:halphai}, respectively.
Writing out Eq.~\ref{eq:halphai} explicitly in terms of matrices, one then obtains 
\begin{align}
\begin{array}{l}
	v_u
\begin{pmatrix}
\frac{h_{e1}}{\sqrt{M_1}}    & \frac{h_{e2}}{\sqrt{M_2}} \\
\frac{h_{\mu1}}{\sqrt{M_1}}  & \frac{h_{\mu2}}{\sqrt{M_2}} \\
\frac{h_{\tau1}}{\sqrt{M_1}} & \frac{h_{\tau2}}{\sqrt{M_2}}
\end{pmatrix} = i
\begin{pmatrix}
0 & U_{e2}^* \sqrt{m_2} & U_{e3}^* \sqrt{m_3} \\
0 & U_{\mu2}^* \sqrt{m_2} & U_{\mu3}^* \sqrt{m_3} \\
0 & U_{\tau2}^* \sqrt{m_2} & U_{\tau3}^* \sqrt{m_3} 
\end{pmatrix}
\begin{pmatrix}
0 & 0 \\
+ \cos \xi & - \sin \xi \\
\zeta \sin \xi & \zeta \cos \xi 
\end{pmatrix}
	\quad \mathrm{[NO]} \\[4ex]
	v_u
\begin{pmatrix}
\frac{h_{e1}}{\sqrt{M_1}}    & \frac{h_{e2}}{\sqrt{M_2}} \\
\frac{h_{\mu1}}{\sqrt{M_1}}  & \frac{h_{\mu2}}{\sqrt{M_2}} \\
\frac{h_{\tau1}}{\sqrt{M_1}} & \frac{h_{\tau2}}{\sqrt{M_2}}
\end{pmatrix} = i
\begin{pmatrix}
U_{e1}^* \sqrt{m_1} & U_{e2}^* \sqrt{m_2} & 0   \\
U_{\mu1}^* \sqrt{m_1} & U_{\mu2}^* \sqrt{m_2} & 0 \\
U_{\tau1}^* \sqrt{m_1} & U_{\tau2}^* \sqrt{m_2} & 0
\end{pmatrix}
\begin{pmatrix}
+ \cos \xi & - \sin \xi \\
\zeta \sin \xi & \zeta \cos \xi \\
0 & 0
\end{pmatrix} 
	\quad \mathrm{[IO]}
\end{array}.
\end{align}
For the ease of notation, let us introduce the following (dimensionless) quantities:
\begin{align}
\kappa_{\alpha i} \equiv \frac{h_{\alpha i}}{\sqrt{M_i / v_u}} \,, \quad
V_{\alpha i} \equiv i\, U_{\alpha i}^* \sqrt{m_i / v_u} \,.
\end{align}
The Casas-Ibarra parametrisation can then be written in the following compact form,
\begin{align}
\begin{pmatrix}
\kappa_{\alpha 1}\left(\xi\right) \\
\kappa_{\alpha 2}\left(\xi\right)
\end{pmatrix} =
\begin{pmatrix}
+\cos\xi & \zeta \sin \xi \\
-\sin \xi & \zeta \cos \xi
\end{pmatrix}
\begin{pmatrix}
V_{\alpha k} \\
V_{\alpha l}
\end{pmatrix} \,,
\end{align}
where $(k, l) = (2, 3)$ in the NH case and $(k, l) = (1, 2)$ in the IH case.

Next, let us distinguish explicitly between the two branches $\zeta = +1$ and $\zeta = -1$.
This provides us with two sets of Yukawa couplings, $\kappa_{\alpha i}^{(+)}$ and $\kappa_{\alpha i}^{(-)}$,
\begin{align}
\begin{pmatrix}
\kappa_{\alpha 1}^{(\pm)}\left(\xi\right) \\
\kappa_{\alpha 2}^{(\pm)}\left(\xi\right)
\end{pmatrix} =
\begin{pmatrix}
+ V_{\alpha k}\, \cos\xi \pm V_{\alpha l}\, \sin \xi \\
- V_{\alpha k}\, \sin \xi \pm V_{\alpha l}\, \cos\xi
\end{pmatrix} \,.
\end{align}
The couplings in the negative branch are related to those in the positive branch
as follows,
\begin{align}
\kappa_{\alpha 1}^{(-)}\left(\xi\right) = 
\kappa_{\alpha 1}^{(+)}\left(-\xi\right) \,, \quad
\kappa_{\alpha 2}^{(-)}\left(\xi\right) = 
- \kappa_{\alpha 2}^{(+)}\left(-\xi\right) \,.
\end{align}
Meanwhile, the couplings in both branches exhibit the following ``periodicity'',
\begin{align}
\kappa_{\alpha 1}^{(\pm)}\left(\xi + \frac{\pi}{2}\right) = \kappa_{\alpha 2}^{(\pm)}\left(\xi\right) \,, \quad
\kappa_{\alpha 2}^{(\pm)}\left(\xi + \frac{\pi}{2}\right) = - \kappa_{\alpha 1}^{(\pm)}\left(\xi\right) \,.
\end{align}
Combining these two properties of the couplings $\kappa_{\alpha i}^{(\pm)}$, we find the following relation,
\begin{align}
\kappa_{\alpha 1}^{(-)}\left(\xi\right) = 
- \kappa_{\alpha 2}^{(+)}\left(\frac{\pi}{2} - \xi\right)
\,, \quad
\kappa_{\alpha 2}^{(-)}\left(\xi\right) =
- \kappa_{\alpha 1}^{(+)}\left(\frac{\pi}{2} - \xi\right) \,.
\end{align}
This illustrates that the couplings in the negative branch follow from the couplings in the positive branch after performing three operations:
(i) flip the sign of all couplings,
$\kappa_{\alpha i} \rightarrow - \kappa_{\alpha i}$,
(ii) exchange the two columns in the neutrino Yukawa matrix,
$\kappa_{\alpha 1} \leftrightarrow \kappa_{\alpha 2}$, and
(iii) perform a reflection in the complex $\xi$ plane about the
$\textrm{Re}\left[\xi\right] = \pi/4$ axis.

None of these steps has the potential to affect our conclusions regarding the final baryon asymmetry as a function of $\xi$.
First of all, step (i) has no consequences for our analysis as we never encounter any solitary single powers of Yukawa couplings $h_{\alpha i}$.
All Yukawa couplings always come at least in pairs, as in quantities such  as $h^\dagger h$.
Other quantities, such as $\mathcal{H}_i$, even contain only fourth powers of Yukawa couplings.
Flipping the sign of all Yukawa couplings at once, therefore, has no effect on the baryon asymmetry.
Step (ii) is irrelevant as we are restricting ourselves to the nearly-degenerate case, $M_1 \simeq M_2$. 
From the perspective of the charged-lepton fields, $\ell_\alpha$, the two neutrinos $N_1$ and $N_2$ are therefore, in a sense, indistinguishable. 
Exchanging the Yukawa couplings, $h_{\alpha 1} \leftrightarrow h_{\alpha 2}$, has no effect except for some sign changes here and there.
However, as we are mainly interested in the absolute value of the baryon asymmetry, $|\eta_B|$, sign changes in $\eta_B$ do not really bother us.
Once we know where in parameter space we can find the right baryon asymmetry with a negative sign, we are immediately able to obtain the right baryon asymmetry with a positive sign by flipping the sign of $\textrm{Re}\left[\xi\right]$. 
Finally, step (iii) is irrelevant as well as our results for $\eta_B$ turn out to be mirror-symmetric w.r.t.\ reflections about the $\textrm{Re}\left[\xi\right] = \pi/4$ axis.%
\footnote{We note that, even if this was not the case, scanning $\textrm{Re}\left[\xi\right]$ over the entire interval from $0$ to $\pi/2$ would be enough to fully capture all relevant Yukawa configurations in both branches.}
This is evident from Fig.~\ref{fig:eta_vs_rexi} and follows from the fact that the final asymmetry is proportional to $\sin\left(2\,\textrm{Re}\left[\xi\right]\right)$ (see the discussion below Fig.~\ref{fig:eta_vs_rexi}).

All in all, we conclude that the couplings in the negative branch, $\kappa_{\alpha i}^{(-)}$, lead to the same qualitative results for the baryon asymmetry as the couplings in the positive branch, $\kappa_{\alpha i}^{(+)}$. 
This justifies our decision to focus on only one branch of the Casas-Ibarra parametrisation.
Our numerical code confirms that $\kappa_{\alpha i}^{(+)}$ and $\kappa_{\alpha i}^{(-)}$ lead to the same results for $\eta_B$.

\FloatBarrier

\end{document}